\documentclass[twocolumn,superscriptaddress,letter]{revtex4}
\usepackage{amssymb}
\usepackage{color}
\usepackage{graphicx}
\usepackage{dcolumn}
\usepackage{subfigure}
\usepackage{bm}
\usepackage[header,title,page,titletoc]{appendix}
\usepackage{amsmath}
\usepackage{bbm}
\usepackage{amsfonts}%
\newcommand{\beq}{\begin{eqnarray}}
\newcommand{\eeq}{\end{eqnarray}}
\newcommand{\be}{\begin{equation}}
\newcommand{\ee}{\end{equation}}
\newcommand{\bw}{\begin{widetext}}
\newcommand{\ew}{\end{widetext}}
\newcommand{\ba}{\begin{array}}
\newcommand{\ea}{\end{array}}

\begin{document}
\title{Radiative transition decay width of $\psi_2(3823)\rightarrow\gamma\chi_{c1}$ from lattice QCD}

%
%

\author{Ning Li}
\email{lining@xatu.edu.cn}
\affiliation{\small School of Sciences, Xi'an Technological University, Xi'an 710032, People's Republic of China}

\author{Yan Gao}
\affiliation{\small School of Sciences, Xi'an Technological University, Xi'an 710032, People's Republic of China}

\author{Feiyu Chen}
\affiliation{\small Institute of High Energy Physics, Chinese Academy of Sciences, Beijing 100049, People's Republic of China}
\affiliation{\small School of Physical Sciences, University of Chinese Academy of Sciences, Beijing 100049, People's Republic of China}

\author{\small Ying Chen}
\affiliation{\small Institute of High Energy Physics, Chinese Academy of Sciences, Beijing 100049, People's Republic of China}
\affiliation{\small School of Physical Sciences, University of Chinese Academy of Sciences, Beijing 100049, People's Republic of China}

\author{Xiangyu Jiang}
\affiliation{\small Institute of High Energy Physics, Chinese Academy of Sciences, Beijing 100049, People's Republic of China}
\affiliation{\small School of Physical Sciences, University of Chinese Academy of Sciences, Beijing 100049, People's Republic of China}

\author{\small Chunjiang Shi}
\affiliation{\small Institute of High Energy Physics, Chinese Academy of Sciences, Beijing 100049, People's Republic of China}
\affiliation{\small School of Physical Sciences, University of Chinese Academy of Sciences, Beijing 100049, People's Republic of China}

\author{Wei Sun}
\affiliation{\small Institute of High Energy Physics, Chinese Academy of Sciences, Beijing 100049, People's Republic of China}

\begin{abstract}
We present an exploratory $N_f=2$ lattice QCD study of $\psi_2(3823)\to \gamma \chi_{c1}$ at a pion mass $m_{\pi}\approx 350$~MeV. The related two-point and three-piont functions are calculated using the distillation method. The electromagnetic multipole form factor $\hat{V}(0)=2.083(11)$ for $J/\psi\to\gamma \eta_c$ is consistent with previous lattice results, the form factors $\hat{E}_1(0)$, $\hat{M}_2(0)$ and $\hat{E}_3(0)$ for $\Gamma(\chi_{c2}\to\gamma J/\psi)$ have the same hierarchy as that derived from experiments and the predicted decay width $\Gamma(\chi_{c2}\to\gamma J/\psi)=368(5)~\text{keV}$ is in excellent agreement with the PDG value $374(10)~\text{keV}$ and previous lattice QCD results in the quenched approximation. The same strategy is applied to the study of the process $\psi_2(3823)\to \gamma \chi_{c1}$ and the partial decay width is predicted to be $337(27)~\text{keV}$. According to the BESIII constraints on the $\psi_2(3823)$ decay channels and some phenomenological results, we estimate the total width $\Gamma(\psi_2(3823))=520(100)~\text{keV}$.

%
%

\end{abstract}



\maketitle
\section{INTRODUCTION}
Charmonium states are usually thought of as the bound states of charm quark and antiquark ($c\bar{c}$) in the conventional quark model.
Since the charm quark is relatively heavy, a non-relativistic description of the internal structure of charmonium is acceptable to some extent,
especially for the low-lying states. In the non-relativistic potential model, a charmonium state can be assigned to a $n^{2S+1}L_J$ state,
where $n$, $S$ and $L$ are the radial quantum number, the total spin of the $c\bar{c}$ pair, and the orbital angular momentum, respectively.
Consequently, it gives the $J^{PC}$ quantum number of the state. For $n=1$, the $S$-wave charmonium ($J/\psi$ and $\eta_c$) and the $P$-wave charmonium
($h_c$ and $\chi_{c0,1,2}$) has been well established, but the $D$-wave super-multiplet $(1^{1}D_2, 1^{3}D_{1,2,3}$) is not complete yet.
Experimentally, apart from the vector charmonium $\psi(3770)$ that is assigned to be (predominantly) the $1^{3}D_1$ state, other $1D$ charmonium have escaped from the
experimental search for a long time. In 2013, the Belle Collaboration reported the first evidence for a $2^{--}$ charmonium-like state $X(3823)$ of a mass $3823.1\pm 1.8\pm 0.7$~$\mathrm{MeV}$ in the $\chi_{c1}\gamma$ invariant mass spectrum of the
decay processes $B\rightarrow\chi_{c1}\gamma{K}$~\cite{Belle:2013ewt}. In 2015, the BESIII collaboration also observed $X(3823)$ in the $\gamma\chi_{c1}$ system with a statistical significance of $6.2\sigma$ in the process $e^+e^-\to \pi^+\pi^-\chi_{c1}\gamma$~\cite{BESIII:2015iqd}. The mass of $X(3823)$ is measured to be $3823.7\pm 1.3\pm 0.7~\mathrm{MeV}$, which is consistent with that measured by Belle and confirms the existence of $X(3823)$. The properties of $X(3823)$, such as its mass and decay modes $\chi_{c1,2}\gamma$~\cite{BESIII:2021qmo}, are consistent with the theoretical expectations for those of the $1D$ state $1^{3}D_2$. Now $X(3823)$ is named by $\psi_2(3823)$ in PDG~\cite{ParticleDataGroup:2022pth}. The observation of $\psi_2(3823)$ in the process $e^+e^-\to \pi^0\pi^0 \psi_2(3823)$ by BESIII~\cite{BESIII:2021qmo,BESIII:2022cyq} provides a further support of its quantum number $J^{PC}=2^{--}$. Recently, the LHCb collaboration observed a new charmonium state $X(3842)$ near the $D\bar{D}$ threshold using proton-proton collision data~\cite{LHCb:2019lnr}. Its mass $m_{X(3842)}=3842.72\pm0.16\pm 0.12~\mathrm{MeV}$ and  the very small width $\Gamma_{X(3842)}=2.79\pm0.51\pm0.35~\mathrm{MeV}$ suggest $X(3842)$ to be a candidate for the $1^3D_3$ charmonium state (named as $\psi_3$ in PDG). Thus the $1D$ spin triplet is in space, while the spin singlet $1D$ state $\eta_{c2}$ is still missing.

The width of $\psi_2(3823)$ is expected to be very small since it lies a little higher than the $D\bar{D}$ threshold but lower than $D\bar{D}^*$ and $D^*\bar{D}^*$ threshold. It cannot decay into $D\bar{D}$ owing to the angular momentum conservation.  Thus, its major decay modes should be radiative and hadronic transitions into other charmonium states. Phenomenological studies predict the partial widths $\Gamma(\psi_2\to\gamma\chi_{c1})\sim 200-300~\text{keV}$ and $\Gamma(\psi_2\to\gamma\chi_{c2})\sim 60~\text{keV}$~\cite{Ebert:2002pp,Barnes:2005pb},
and $\Gamma(\psi_2\to J/\psi \pi\pi)\sim 160~\text{keV}$~\cite{Wang:2015xsa} (However, BESIII gives the upper limits $\Gamma(\psi_2(3823)\to \pi^+\pi^- J/\psi)/\Gamma(\psi_2(3823)\to \gamma \chi_{c1})<0.06$ and $\Gamma(\psi_2(3823)\to \pi^0\pi^0 J/\psi)/\Gamma(\psi_2(3823)\to \gamma \chi_{c1})<0.11$, which are in striking contrast to the theoretical expectation). This indicates $\psi_2\to \gamma\chi_{c1}$ might be the most important decay channel. Experimentally, LHCb gives an upper bound $\Gamma_{\psi_2}<5.2~\text{MeV}$~\cite{LHCb:2020fvo}, while a recent BESIII measurement decreases this limit to be $\Gamma_{\psi_2}<2.9~\text{MeV}$ at the 90\% confidence level~\cite{BESIII:2022yga}. So reliable determination of the partial width $\psi_2(3823)\to \gamma \chi_{c1}$ is very helpful to estimate the total width of $\psi_2(3823)$.


A first-principles calculation of $\psi_2(3823)$ decays is desired in two folds. Firstly, charmonium states are located at the intermediate energy scale of QCD, where both perturbative and nonperturbative physics are present, and charmonium states are considered as an ideal test ground for quantum chromodynamics (QCD). Secondly, the comparison of the quark model predictions and the first principle calculation can indicate to what extent the quark model describes the properties of charmonium. The numerical Lattice QCD calculation is known as an \textit{ab initio} approach to solve the low energy problems of QCD, and has been extensively applied to the study of radiative transition between various  charmonium~\cite{Dudek:2006ej,Dudek:2006ut,Dudek:2009kk,Chen:2011kpa,Yang:2012mya,Gui:2012gx,Donald:2012ga,Yang:2013xba,Becirevic:2014rda,Gui:2019dtm,Meng:2019lkt,Liu:2020qfz,Li:2021gze,Jiang:2022gnd,Chen:2022isv,Delaney:2023fsc,Colquhoun:2023sti}.

In this simulation, we calculate the radiative transition decay width of $\psi_2\rightarrow\chi_{c1}\gamma$ in the framework of $N_f=2$ lattice QCD. We compute related two-point and three-point correlation functions by the implementation of the distillation method~\cite{Peardon:2009gh,Shultz:2015pfa,Delaney:2023fsc}. This smearing technique helps us use optimized operators of definite momentum at both source and sink as well as insert a vector current operator of definite momentum. Therefore, it has efficiently decreased the errors of physical quantities extracted from the correlation functions.
As a calibration of possible systematic uncertainties with our lattice setup, we also calculate the radiative transition decay width of $J/\psi\rightarrow\eta_c\gamma$ and $\chi_{c2}\rightarrow{J/\psi\gamma}$ and compare them with previous lattice results and experimental values.

This paper is organized as follows. In Sect.~\ref{lattice setup}, the strategies for computing form factors and the decay width of radiative transition have been briefly reviewed. Sect.~\ref{Simulation details} is composed of two parts. In Sect.~\ref{charmonium31}, we briefly introduce the distillation method for computing two-point correlations, and the mass spectrum for $\eta_c$, $J/\psi$, $\chi_{c1}$, $\chi_{c2}$ and $\psi(3823)$ are listed. In Sect.~\ref{charmonium32}, we briefly introduce the distillation method for computing three-point correlation functions. Sect.~\ref{Form Factor of charmonium} is divided into three parts. In Sect.~\ref{results1}, the numerical results for $J/\psi\rightarrow\eta_c\gamma$ are listed. In Sect.~\ref{resluts2}, the numerical results for $\chi_{c2}\rightarrow{J/\psi\gamma}$ are listed. In Sect.~\ref{resluts3}, the numerical results for
$\psi_2\rightarrow\chi_{c1}\gamma$ are listed. In Sect.~\ref{Discussion}, we make brief discussions and draw conclusions.

\section{Formalism}
\label{lattice setup}
For a radiative transition process $i(p_i)\to \gamma f(p_f)$, the partial decay width can be expressed in terms of
the electromagnetic multipole form factors $F_k(Q^2)$ at $Q^2=0$, namely,
\beq\label{eq:decay-width}
\Gamma(i\rightarrow{f}\gamma)=\frac{1}{2J_i+1}\alpha\frac{|\mathbf{q}|}{m_i^2}\sum_k|F_k(0)|^2,
\label{decay width}
\eeq
where $\alpha=1/137$ is the fine structure constant at the charm quark scale, $\mathbf{q}$ is the momentum of the photon in the final state with $|\mathbf{q}|=\frac{(m_i^2-m_f^2)}{2m_i}$ and $Q^2=-q^2=(p_i-p_f)^2$. The multipole form factors $F_k(Q^2)$ are encoded in the matrix element of the electromagnetic current $J_\mu^{\text{em}}(0)$ between the initial and final hadron states, namely,
\begin{eqnarray}\label{eq:multipole}
   && \langle f,p_f,r_f|J_\mu^\mathrm{em}(0)|i,p_i,r_i\rangle\nonumber\\
   && \equiv \sum\limits_k \alpha^k_\mu(p_i,p_f,\epsilon^{(r_f),*},\epsilon^{(r_i)})F_k(Q^2),
\end{eqnarray}
where $\epsilon^{(r)}$ refers to the polarization vectors (tensors) of the initial and the final hadron states, $\alpha_k$ are known functions of $p_i, p_f, \epsilon^{(r_{i,f})}$ that are determined through the multipole decomposition~\cite{Dudek:2006ej,Dudek:2009kk,Yang:2012mya}.

The matrix element on the left hand side can be extracted from the following three-point correlation functions with an insertion of the local current $J_\mu^\mathrm{em}(x)$, i.e.,
\beq\label{eq:three-point}
G_{f\mu i}(t_f,t;\mathbf{p}_f,\mathbf{p}_i)&=&\sum_{\mathbf{x},\mathbf{y}}e^{-i\mathbf{p}_f\cdot{\mathbf{x}}}
e^{i\mathbf{q}\cdot{\mathbf{y}}}\nonumber\\
& &\times\langle\Omega|T\mathcal{O}_f(t_f,\mathbf{x})J^\mathrm{em}_{\mu}(t,\mathbf{y})\nonumber\\
& &\times\mathcal{O}^{\dag}_i(0,\mathbf{0})|\Omega\rangle,
\label{three point function}
\eeq
where $\mathcal{O}_f(t_f,\mathbf{x})$ and $\mathcal{O}^{\dag}_i(0,\mathbf{0})$ are interpolating operators for the final and the initial hadron states, respectively, $\mathbf{q}=\mathbf{p}_i-\mathbf{p}_f$ is the momentum of the (virtual) photon, and $J^{\text{em}}_{\mu}$ is the electromagnetic vector current whose explicit form is
\beq\label{eq:current}
\label{conseved current1}
J^\text{em}_{\mu}(x)&=&\sum_q Q_q\bar{\psi_q}(x)\gamma_{\mu}\psi_q(x)
\\\nonumber&\to& \frac{2e}{3} \bar{c}(x) \gamma_\mu c(x)\\\nonumber
&=&\frac{2e}{3}J_{\mu}
\eeq
with $q$ referring to $u,d,s,c,b$ quark flavors. For charmonium radiative decays, since $u,d,s,b$ quarks contribute to $G_{f\mu i}(t_2,t;\mathbf{p_2},\mathbf{p_1})$ through disconnected quark diagrams, which are suppressed by OZI rules, we only consider the electromagnetic current of charm quark with $Q_c=2e/3 $ in the practical calculation.

After inserting a complete set of states between the electromagnetic vector current and the
interpolating operators, Eq.~(\ref{three point function}) has the following asymptotic form in the $t_f\gg{t}\gg1$ limit,
\beq
G_{f\mu i}(t_f,t;\mathbf{p}_f,\mathbf{p}_i) &\overset{t_f\gg{t}\gg1}\longrightarrow&\frac{e^{-E_ft_f}e^{-(E_i-E_f)t}}{4E_i(\mathbf{p}_i)E_f(\mathbf{p}_f)}\nonumber\\
& &\times\langle\Omega|\mathcal{O}_f|f(\mathbf{p}_f)\rangle\langle{i(\mathbf{p}_i)}|\mathcal{O}_i^{\dag}|\Omega\rangle\nonumber\\
& &\times\langle{f(\mathbf{p}_f)}|J^{\text{em}}_{\mu}(0)|i(\mathbf{p}_i)\rangle.
\label{three point function_1}
\eeq
In order to extract the matrix element $\langle{f}|J^\text{em}_{\mu}(0)|i\rangle$, we
need to the energies and spectral weight of the final and the initial states, i.e., $E_i$, $E_f$, $\langle\Omega|\mathcal{O}_f|f(\mathbf{p}_f)\rangle$,
and $\langle{i(\mathbf{p}_i)}|\mathcal{O}_i^{\dag}|\Omega\rangle$. They can be determined from the two-point correlation function,
\begin{eqnarray}\label{eq:two-point}
C(t,\mathbf{p}) &=& \sum_{\mathbf{x}} e^{ -i\mathbf{p} \cdot \mathbf{x} }\langle\Omega|\mathcal{O}(t,\mathbf{x})
                    \mathcal{O}^{\dag}(0,\mathbf{0})|\Omega\rangle\nonumber\\
                &=&\sum_{n} \frac{ |\langle \Omega|\mathcal{O}(0)|n,\mathbf{p}\rangle|^2}{2E_n(\mathbf{p})}e^{-E_n(\mathbf{p})t}\nonumber\\
                & \to & \frac{|Z(\mathbf{p})|^2} {2E(\mathbf{p})} e^{-E(\mathbf{p})t}~~~(t\to \infty),
\end{eqnarray}
where $E(\mathbf{p})$ is the energy of the ground state $|1,\mathbf{p}\rangle$ and $Z(\mathbf{p})=\langle \Omega|\mathcal{O}(0)|1,\mathbf{p}\rangle$ is defined.
So the key problem in this work is to calculate the two-point and three-point functions, from which
the transition amplitude can be derived. It should be notified that the polarizations of $J\ne 0$ particles are not spelt out explicitly in the discussion above for simplicity, but are taken into account the in the concrete calculations.

\begin{table}[t]
    \renewcommand\arraystretch{1.5}
    \caption{Parameters of the gauge ensemble.}
    \label{tab:config}
    \begin{ruledtabular}
        \begin{tabular}{lllllc}
            $L^3\times T$     & $\beta$ & $a_t^{-1}$(GeV) & $\xi$      & $m_\pi$(MeV) & $N_\mathrm{cfg}$ \\\hline
            $16^3 \times 128$ & 2.0     & $6.894(51)$     & $\sim 5.3$ & $348.5(1.0)$ & $689$           \\
        \end{tabular}
    \end{ruledtabular}
\end{table}

\section{Simulation details}\label{Simulation details}

 We use a subset of the $N_f=2$ gauge ensemble generated on an $L^3\times T=16^3\times 128$ anisotropic lattice with the anisotropy parameter $\xi=a_s/a_t=5.3$ ($a_s$ and $a_t$ are the spatial and temporal lattice spacing, respectively)~\cite{Jiang:2022ffl}. The sea quark mass is tuned to give the pion mass $m_\pi\approx 350$ MeV. The parameters of the gauge ensemble are listed in Table~\ref{tab:config}. For the valence charm quark, we adopt the clover fermion action in Ref.~\cite{CLQCD:2009nvn} and the charm quark mass parameter is set by $(m_{\eta_c}+3m_{J/\psi})/4=3069$ MeV. For each source time slice $\tau\in [0,T-1]$ on each gauge configuration, the perambulators of charm quark are calculated in the Laplacian Heaviside (LH) subspace spanned by $N_{\mathrm{vec}}=50$ eigenvectors with lowest eigenvalues.
\begin{table*}[t]
\centering
\caption{The interpolating operators~\cite{Dudek:2007wv} and masses of charmonium states involved in this work. The PDG masses values~\cite{ParticleDataGroup:2022pth} of these states are also presented for comparison.}
 \label{tab:gamma mtrix}
 \begin{ruledtabular}
 \begin{tabular}{cccccc}
 Meson       &$\eta_c$       &$J/\psi$      &$\chi_{c1}$           &$\chi_{c2}$   &$\psi_2$ \\
\hline
$\Gamma$      &$\gamma_{5}$           &$\gamma_{i}$         &$\gamma_{i}\gamma_5$          &$|\epsilon_{ijk}|\gamma_j\triangledown_{k}(\mathcal{Q}_{ijk}\gamma_j\triangledown_{k})$ &$|\epsilon_{ijk}|\gamma_5\gamma_j\triangledown_{k}(\mathcal{Q}_{ijk}\gamma_5\gamma_j\triangledown_{k})$\\
$m(\mathrm{MeV})$                                      &2976.8(0.4)      &3099.9(0.4)      &3563.1(1.6)   &3610.8(1.7)           &3907.5(7.6)\\
$m(\mathrm{MeV)(PDG})$~\cite{ParticleDataGroup:2022pth} &2983.9(0.4)        &3096.900(0.006)      &3510.67(0.05)     &3556.17(0.07)            &3823.7(0.5)\\
 \end{tabular}
 \end{ruledtabular}
 \end{table*}
\subsection{Charmonium spectrum}\label{charmonium31}

\begin{figure}[t]
  {\includegraphics[width=8cm]{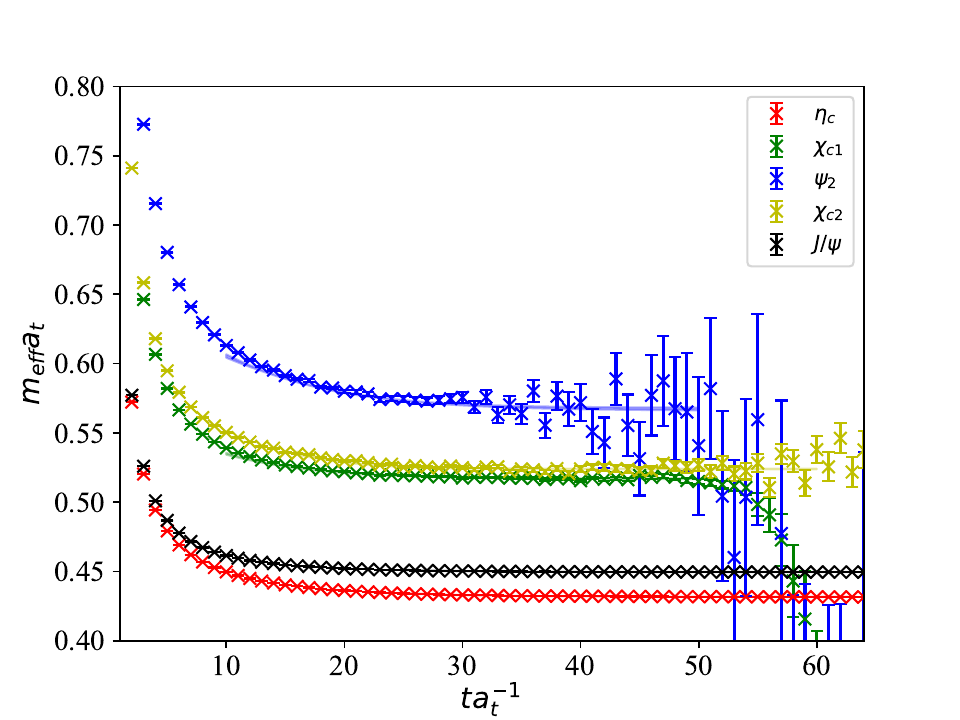}}
\caption{Charmonium effective mass plateaus.
From lower to higher values, the plateau corresponds to the
charmonium state $\eta_c$, $J/\psi$, $\chi_{c1}$, $\chi_{c2}$, $\psi_{2}$, respectively.}
\label{fig:meson_mass}
\end{figure}

In this section, we introduce briefly the distillation method to compute two point correlation function~\cite{Peardon:2009gh}.
The distillation method provides automatically the Laplacian Heaviside (LH) smearing scheme for quark fields. The LH smeared charm quark field on each time slice $t$ is defined as
\begin{equation}
    c^{(s)}(\mathbf{x},t)=\sum\limits_\mathbf{y} \square_{\mathbf{x},\mathbf{y}}(t) c(\mathbf{y},t),
\end{equation}
with the smearing function $\square_{\mathbf{x},\mathbf{y}}(t)$ being defined by the eigenvectors $\{\xi_{\mathbf{x}}^{(n)}(t),n=1,2,\ldots,N_\text{vec}\}$ that span the LH subspace, namely,
\beq
\square_{\mathbf{x}\mathbf{y}}(t)=\sum_{n=1}^{N}\xi_{\mathbf{x}}^{(n)}(t)\xi_{\mathbf{y}}^{(n)\dagger}(t).
\eeq
Subsequently, each interpolation operator $\mathcal{O}$ in Eq.~(\ref{eq:three-point}) and Eq.~(\ref{eq:two-point}) is built in terms of $c^{(s)}$
\begin{equation}\label{eq:operators}
\mathcal{O}(t,\mathbf{x})=\sum\limits_\mathbf{y} \bar{c}^{(s)}(t,\mathbf{x})\Gamma(\mathbf{x},\mathbf{y};t) c^{(s)}(t,\mathbf{x}),
\end{equation}
where $\Gamma(\mathbf{x},\mathbf{y};t)$ is a specific combination of $\gamma$ matrices and the discretized covariant derivatives and dictates the quantum number of the operators (The $\Gamma$'s for the charmonium states involved in this work are listed in Table~\ref{tab:gamma mtrix}).
A normal Fourier transformation can project out the operator that annihilates a charmonium state with a definite spatial momentum $\mathbf{p}$
\beq
\mathcal{O}(t,\mathbf{p})&=&\sum\limits_{\mathbf{y}} e^{-i\mathbf{p}\cdot\mathbf{y}} \mathcal{O}(t,\mathbf{x})\nonumber\\
&\equiv& [\bar{c}_{\mathbf{x}}\square_{\mathbf{x}\mathbf{y}}e^{-i\mathbf{p}\cdot\mathbf{y}}\Gamma_{\mathbf{y}\mathbf{z}}\square_{\mathbf{z}\mathbf{w}} c_{\mathbf{w}}](t)\nonumber\\
&\equiv& [\bar{c}\square \Gamma(\mathbf{p})\square c](t)
\eeq
where the subscripts $\mathbf{x,y,z,w}$ in the second row means that the spatial coordinates are viewed as matrix indices with the duplicated subscripts being summed implicitly, and $\Gamma(t,\mathbf{p})$ in the third row is $[\Gamma(p)]_\mathbf{xy}(t)=e^{-i\mathbf{p}\cdot\mathbf{x}}\Gamma_\mathbf{xy}(t)$. The two-point correlation function can be expressed as
\beq
C(t,\mathbf{p})&=&\langle\Omega|[\bar{c}\square\Gamma(\mathbf{p})\square c](t)[\bar{c}\square\Gamma(\mathbf{p})\square c](0)|\Omega\rangle\nonumber\\
&=&\tau_{nm}(0,t)\Phi_{mp}^\Gamma (t,\mathbf{p})\tau_{pq}(t,0)\Phi^\Gamma_{qn}(0,\mathbf{p}),
\eeq
where $\tau_{pq}(t,0)=\xi_{p}^\dagger(t)M^{-1}(t,0)\xi_{q}(0)$ is the perambulators that is obtained by inverting the Dirac matrix $M$
on sources $\xi_{q}(0) \{q=1,\cdots,N_\mathrm{vec}\}$. $\Phi_{mp}^\Gamma(t,\mathbf{p})=[\xi_{m}^\dagger\Gamma(\mathbf{p})\xi_{p}](t)$ is the elemental that reflects the structure of the corresponding operator.
In this study, we calculate the spectrum of charmonium states $\eta_c$, $J/\psi$, $\chi_{c1}$, $\chi_{c2}$, $\psi_2$). The mass values for
these meson is listed in Table~\ref{tab:gamma mtrix}, where also listed are the $\Gamma$ operators in Eq.~(\ref{eq:operators}) for the charmonium states involved in this work. Figure~\ref{fig:meson_mass} shows the effective mass functions $m_\text{eff}a_t=\ln \frac{C(t,\mathbf{0})}{C(t+a_t,\mathbf{0})}$ of the correlation functions of these charmonium states.
\begin{figure}[t]
    \includegraphics[width=0.8\linewidth]{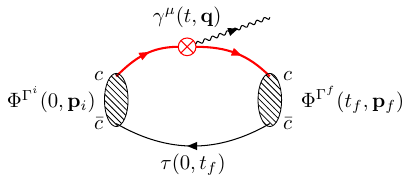}
\caption{The schematic diagram of the calculation three-point function using the distillation method. The filled black line is the perambulator $\tau(0,t_f)$ of the charm quark. The red line with the vector current insertion is the generalized perambulator $\mathcal{G}_{pq\mu}(t_f,t,0)$ in Eq.~(\ref{eq:generalized}). The hatched ellipses indicates the wave functions $\Phi^{i,f}$ of the initial and final charmonia.}
\label{fig:schematic}
\end{figure}

\subsection{Three-point functions}\label{charmonium32}

Since the operators $\mathcal{O}_{i,f}$ are constructed in terms of the LH smeared quark fields (see Eq.~(\ref{eq:operators})), the Wick's contraction of the three-point correlation function in Eq.~(\ref{eq:three-point}) results in the explicit expression for a given $t_f$
\beq\label{eq:generalized}
&&G_{f\mu i}(t_f,t;\mathbf{p}_f, \mathbf{p}_i)\nonumber\\
&\to&\langle\Omega|[\bar{c}\square\Gamma^{f}(\mathbf{p}_f)\square c](t_f)J_{\mu}(t)[\bar{c}\square\Gamma^{i}(\mathbf{p}_i)\square c(0)]|\Omega\rangle\nonumber\\
&=&\langle\Omega|[\bar{c}\square\Gamma^{f}(\mathbf{p}_f)\square c](t_f)[\bar{c}\gamma_{\mu} c](t)[\bar{c}\square\Gamma^{i}(\mathbf{p}_i)\square c](0)|\Omega\rangle\nonumber\\
&=&\tau_{nm}(0,t_f)\Phi^{\Gamma^f}_{mp}(t_f,\mathbf{p}_f)\mathcal{G}_{pq\mu}(t_f,t,0)\Phi^{\Gamma^i}_{qn}(0,\mathbf{p}_i).
\eeq
where $\mathcal{G}_{pq\mu}(t_f,t,0)=\xi_p^{\dagger}(t_f)M^{-1}(t_f,t)\Gamma_{\mu}{M^{-1}(t,0)}\xi_q(0)$ is called generalized perambulator. The schematic diagram for the calculation of $G_{f\mu i}(t)$ is shown in Fig.~\ref{fig:schematic}, where the hatched ellipses stands for the wave functions $\Phi^{i,f}$ of the initial and final charmonia, the filled black line is the perambulator $\tau(t_1,t_2)$ of the charm quark, while the red line with the vector current insertion is the generalized perambulator, which are calculated separately owing to the insertion of the local current~\cite{Shultz:2015pfa}.

To reduce the unknown factors in Eq.~(\ref{three point function_1}), the ratio between the three-point function and the two-point function is introduced, i.e.,
\beq
R_{\mu}(t,t_f)&=&\frac{Z_i(\mathbf{p}_i))Z_f(\mathbf{p}_f))G_{f\mu i}(t_f,t;\mathbf{p}_f,\mathbf{p}_i)}{C_f(t_f-t,\mathbf{p}_f)C_i(t,\mathbf{p}_i)}\nonumber\\
&\simeq&\frac{\langle{f(p_f)}|J_{\mu}(0)|i(p_i)\rangle}{4\sqrt{E_f(p_f)E_i(p_i)}}.
\label{eq:ratio}
\eeq
Here, the second line is valid when $t_f\gg{t}\gg1$ and only the ground state dominates.
$C_i$ and $C_f$ are two-point correlation functions of the initial state and the final state respectively.
$G_{f\mu i}(t_f,t;\mathbf{p}_f,\mathbf{p}_i)$ is a three-point correlation function. Since the matrix element $\langle{f(p_f)}|J_{\mu}(0)|i(p_i)\rangle$ is independent of $t$, it can be derived
in the plateau region of $R_{\mu}(t,t_f)$ is independent of $t$ where the ground states of the
initial and final state charmonia dominate the contribution.

After the matrix element is obtained at each value of $Q^2$,
we can use the multipole expansion expression Eq.~(\ref{eq:multipole}) to extract the form factors $F_k(Q^2)$. In order to give a theoretical prediction of the partial decay width using Eq.~(\ref{eq:decay-width}), we need the on-shell form factors $F_k(Q^2=0)$, which can be determined through the interpolation or extrapolation of $F_k(Q^2)$ with respect to $Q^2$. Usually, one can use the quark model-inspired function forms to do the interpolation or
extrapolation (see below), or just use polynomials of $Q^2$ in the neighborhood of $Q^2=0$.

\section{Charmonium radiative transitions}\label{Form Factor of charmonium}

Since we have only one gauge ensemble of a single lattice spacing, a single light quark mass, we first calculate the partial decay widths $\Gamma(J/\psi\to \gamma \eta_c)$ and $\Gamma(\chi_{c2}\to \gamma J/\psi)$. The comparison of our results with those of previous lattice calculations and experimental values is used as a calibration of the possible systematic uncertainties of our lattice setup. Then the similar calculation is applied to the process $\psi_2\to \gamma\chi_{c1}$.

 The continuum current form in Eq.~(\ref{eq:current}) is not conserved on the lattice and should be renormalized. We adopt the strategy used in refs.~\cite{Dudek:2006ej,Yang:2012mya} to determine the renormalization factor $Z_V$. By calculating the relevant electromagnetic form factors of $\eta_c$, we obtain $Z_V^{t}=1.165(3)$ for the temporal component of $J_\mu^\text{em}$ and $Z_V^s=1.118(4)$ for its spatial components~\cite{Jiang:2022gnd,Chen:2022isv}. In this work, only the spatial components of $J_\mu^\text{em}$ are involved in the calculation, and the renormalization constant $Z_V^{s}$ is incorporated implicitly in the current insertion.

 As shown in Fig.~\ref{fig:schematic}, the current insertion to each quark line gives the same result, so we only consider one of the two insertions. On the other hand, the electric charge $Q_c=2e/3$ of charm quark is not included in $J_\mu$ in the practical calculation for simplicity, therefore the form factors $\hat{F}_k(Q^2)$ extracted from three-point functions is related to the original ones $F_k(Q^2)$ in Eq.~(\ref{eq:multipole}) by the convention
 \begin{equation}
     F_k(Q^2)=2\times \frac{2e}{3} \times \hat{F}_k(Q^2).
     \label{eq:relation F}
 \end{equation}
This convention applies to all the form factors considered in this work.

The radiative transitions in this study are all studied in the rest frame of the initial state, namely, the spatial momentum of the initial state is set to be $\mathbf{p}_i=0$ such that $\mathbf{q}=-\mathbf{p}_f$. The momentum mode $\mathbf{n}=(n_1,n_2,n_3)$ of the final state momentum $\mathbf{p}_f=\frac{2\pi}{La_s}\mathbf{n}$ is represented by $(0,0,0), (0,0,1),(0,1,1),(0,0,2),(0,1,2)$ or $(1,1,2)$ in the meaning that all the momentum modes that can be obtained by applying the lattice symmetry operation to each mode $\mathbf{n}$ in the list are sorted in the same mode denoted by $\mathbf{n}$. Obviously, for a specific transition process, the $\mathbf{p}_f$'s in each mode gives the same $Q^2=-(E_f-m_i)^2+\mathbf{p}_f^2$ and the $Q^2$ of different modes are different from each other.

\begin{figure}[t]
  \includegraphics[width=1.0\linewidth]{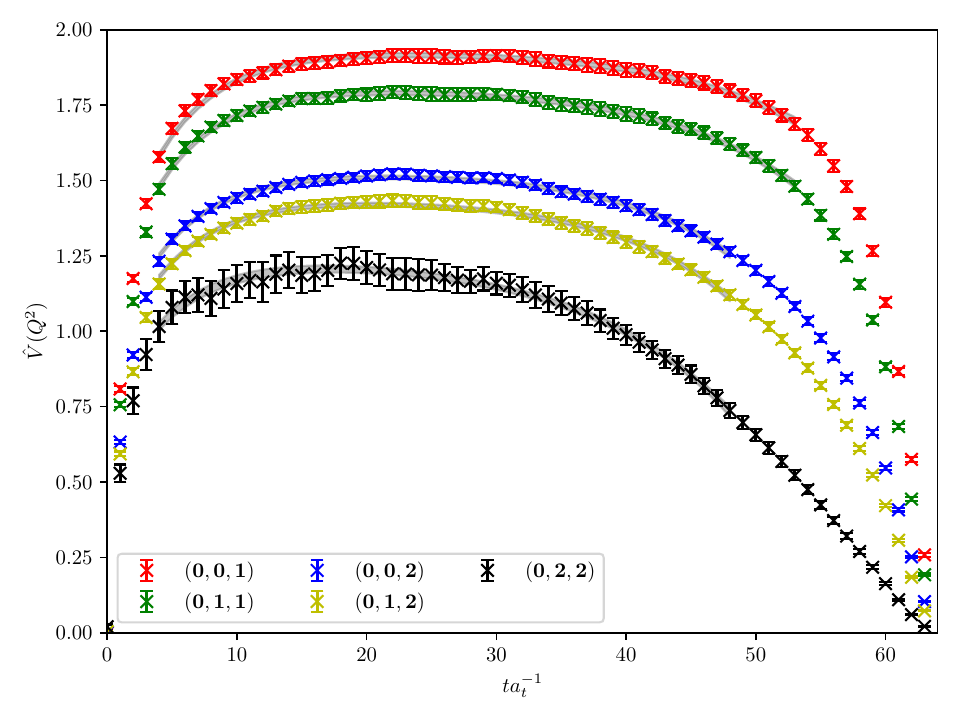}
\caption{The $t$ dependence of $\hat{V}(Q^2)$. The momentum of final particle is $\bf{p_f} = \frac{2\pi}{a_sL}(n_x\ n_y\ n_z)$, the legend denotes explicit value of $(n_x,n_y,n_z)$. The points are lattice data and the shaded bands are the fit results using the function form in Eq.~(\ref{eq:tdependence}).}
\label{fig:fit form factor1}
\end{figure}
\begin{table}[t]
 \centering \caption{The form factor $\hat{V}(Q^2)$ at different $Q^2$.}
 \label{tab:factor V}
 \begin{ruledtabular}
 \begin{tabular}{ccl}
$\mathbf{n}$   &$Q^2(\text{GeV}^2)$         &$\hat{V}(Q^2)$\\\hline
(0,2,2)       &2.04                 &1.288(26)\\
(0,1,2)       &1.29                 &1.499(10)\\
(0,0,2)       &1.04                 &1.512(8)\\
(0,1,1)       &0.52                 &1.784(10)\\
(0,0,1)       &0.25                 &1.910(9)\\\hline
-             & 0                   & 2.083(11)
 \end{tabular}
 \end{ruledtabular}
 \end{table}
\begin{figure}[t]
  {\includegraphics[width=7cm]{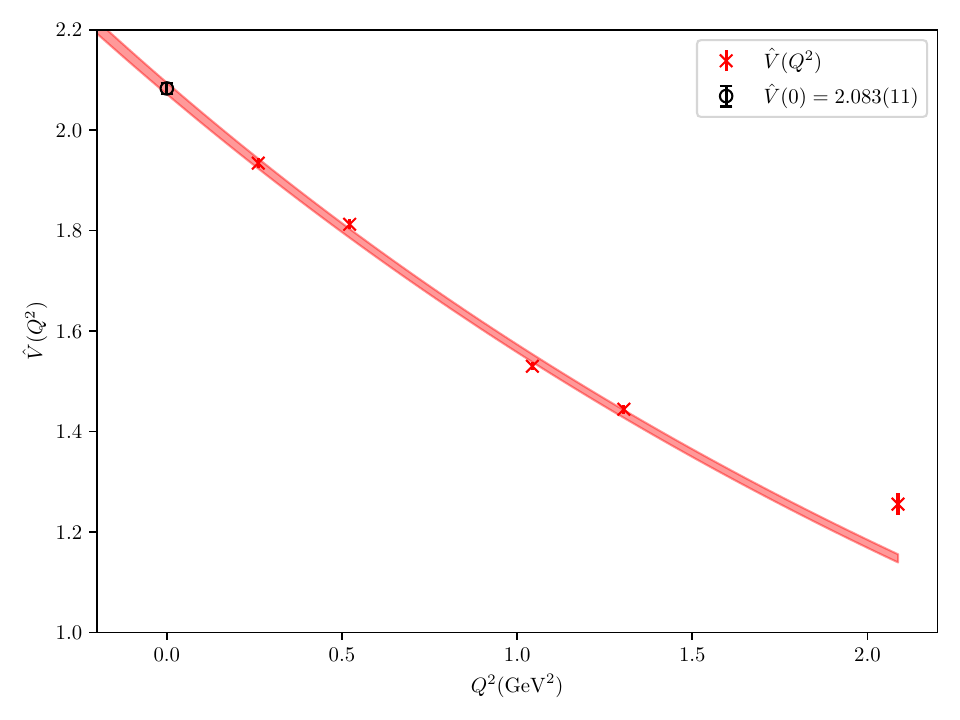}}
\caption{The $Q^2$ extrapolation of $\hat{V}(Q^2)$. The shaded band shows the fit result using Eq.~(\ref{eq:etac-extrapolation}), and the black point is the value of $\hat{V}(Q^2=0)$ through the $Q^2$-extrapolation.}
\label{fig:form factor2}
\end{figure}
\begin{table}[t]
 \centering \caption{The comparison of the form factor $\hat{V}(0)$ in this work with those in previous lattice QCD studies.}
 \label{tab:factor V1}
 \begin{ruledtabular}
 \begin{tabular}{lc}
$\hat{V}(0)$   &reference\\
\hline
1.85(4)         &\cite{Dudek:2006ej}\\
2.01(2)         &\cite{Chen:2011kpa}\\
1.92(3)(2)      &\cite{Becirevic:2012dc}\\
1.90(7)(1)       &\cite{Donald:2012ga}\\
1.83$\sim$2.07        &\cite{Delaney:2023fsc}\\
1.8649(73)       &\cite{Colquhoun:2023zbc}\\
\hline
2.083(11)       &this study\\
 \end{tabular}
 \end{ruledtabular}
 \end{table}
\subsection{$J/\psi\to \gamma\eta_c$}\label{results1}

The transition amplitude for the process
for $J/\psi\rightarrow\eta_c\gamma$ involves only one form factor $V(Q^2)$~\cite{Dudek:2006ej,Chen:2011kpa}

\beq
&&\langle\eta_c(\mathbf{p}_f)|J^{em}_{\mu}|J/\psi(\mathbf{p}_i),r\rangle\nonumber\\
&=&\frac{2V(Q^2)}{m_{\eta_c}+m_{J/\psi}}
\epsilon^{\alpha\mu\beta\gamma}p_{f,\mu}p_{i,\beta}\epsilon_{\gamma}(\mathbf{p}_i,r).
\label{eq:matrix_decomposition}
\eeq
In practice, we derive the transition amplitude using Eq.~(\ref{eq:ratio}) first and then obtain $\hat{V}(Q^2;t,t_f)$ by solving Eq.~(\ref{eq:relation F}) and Eq.~(\ref{eq:matrix_decomposition}) for each momentum $\mathbf{p}_f$ of the final state. For $t_f=64 a_t$, the $t$-dependence of $\hat{V}(Q^2;t,t_f)$ for different values of $Q^2$ is shown in Fig.~\ref{fig:fit form factor1}, where the obvious $t$-dependence
near $t=0$ and $t=t_f$ is attributed to the contamination from higher initial states and higher final states, respectively. Therefore, we use the following
function form,
\beq
\hat{V}(Q^2;t,t_f)=\hat{V}(Q^2)\left(1+\delta_1e^{\Delta_1t}+\delta_2(Q^2)e^{-\Delta_2(t_f-t)}\right)\nonumber\\
\label{eq:tdependence}
\eeq
to fit the data at different $Q^2$ simultaneously. Since we set the initial state to be in its rest frame and let the final state move with a specific momentum, the parameters $\delta_1$ and $\Delta_1$ describe the contribution from the higher initial states and are thereby uniform for all the different values of $Q^2$ involved, while the parameters $\delta_2$ and $\Delta_2$ for the higher final states have $Q^2$-dependence. The fit results are also illustrated by colored bands in Fig.~\ref{fig:fit form factor1}, where one can see the fit form in Eq.~(\ref{eq:tdependence}) describes the data very well. The fitted values of $\hat{V}(Q^2)$ are shown in Table~\ref{tab:factor V}. In order to obtain the on-shell form factor $\hat{V}(Q^2=0)$, which enters the partial decay width as
\beq
\Gamma(J/\psi\to \gamma \eta_c)=\frac{64}{27}\alpha\frac{|\mathbf{q}|^3}{(m_1+m_2)^2}|\hat{V}(0)|^2,
\label{jpsi decay width}
\eeq
we perform a $Q^2$-interpolation using the function form
\beq\label{eq:etac-extrapolation}
\hat{V}(Q^2)=\hat{V}(0)\exp(-\frac{Q^2}{16\beta^2}),
\eeq
inspired by the quark model~\cite{Dudek:2006ej} (As shown in Fig.~\ref{fig:form factor2}). Finally, we get the result
\beq
\hat{V}(0)=2.083(11),~~~\beta=468(3)~\text{MeV}.
\eeq

The value of fitted parameter $\hat{V}(0)$ is consistent with the previous lattice results, as listed in Table~\ref{tab:factor V1}.
By using the experimental values of $m_{J/\psi}$ and $m_{\eta_c}$, we predict the partial width $\Gamma(J/\psi\to \gamma\eta_c)=2.77(3)$~keV, which
is consistent with previous lattice results but is still larger than the PDG average $\Gamma(J/\psi\to \gamma\eta_c)=1.57(37)$~keV~\cite{ParticleDataGroup:2022pth}.

\subsection{$\chi_{c2}\to \gamma J/\psi$}\label{resluts2}

The multipole decomposition of the transition amplitude for the decay $\chi_{c2}\to \gamma J/\psi$ is expressed as
\beq\label{eq:2pp-1mm}
 &&\langle J/\psi(\mathbf{p}_f,r_2)|J^{\text{em}}_{\mu}(0)|\chi_{c2}(\mathbf{p}_i,r_1)\rangle=\alpha_{\mu}^1 E_1(Q^2)\nonumber\\
 &&+\alpha_{\mu}^2 M_2(Q^2)+\alpha_{\mu}^3 E_3(Q^2)+\alpha_{\mu}^4 C_1(Q^2)+\alpha_{\mu}^5 C_2(Q^2),\nonumber\\
\eeq
where $\alpha_{\mu}^k~(k=1,2,\ldots,5)$ are Lorentz covariant kinematic functions of $p_{i,f}$ and polarization vectors of $J/\psi$ and $\chi_{c2}$, whose explicit expressions
can be found in Ref.~\cite{Dudek:2009kk,Yang:2012mya}. It should be noted that the $J=2$ (for example, the spin of $\chi_{c2}$ and $\psi_2$) representation in the continuum breaks into $E$ and $T_2$ irreducible representations of the octahedral group on a finite lattice. It is observed that this breaking effect is negligible as manifested by the nearly degenerate masses of tensor mesons derived from the $E$ operator and $T_2$ operator. Subsequently, the multipole decomposition is performed on the basis of $E\oplus T_2$.

\begin{figure*}[!htbp]
  \centering
  \includegraphics[width=5.5cm]{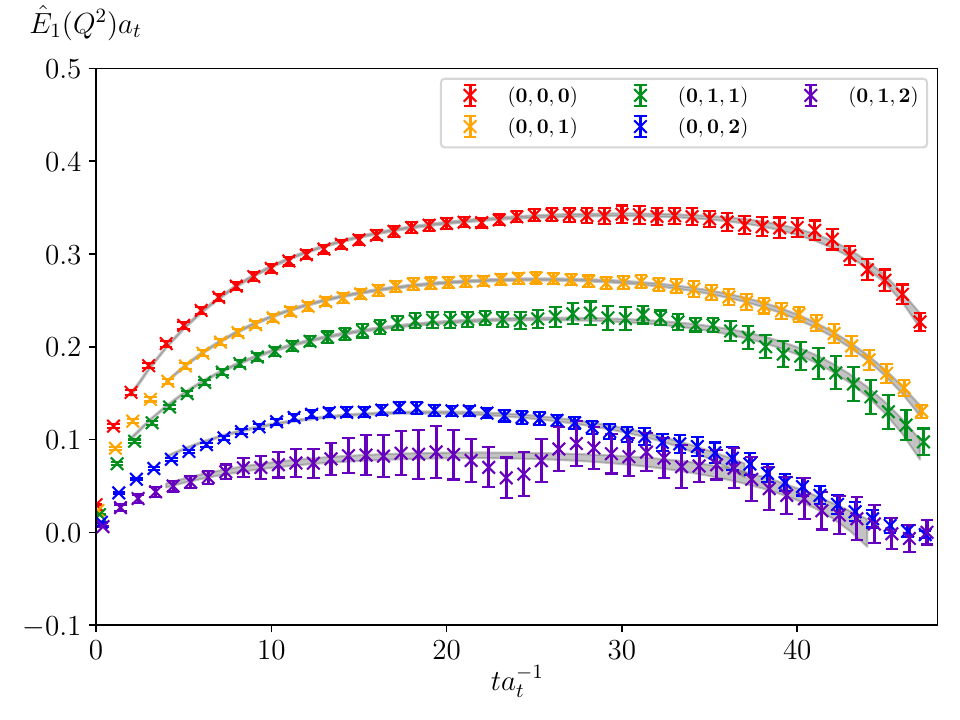}
  \includegraphics[width=5.5cm]{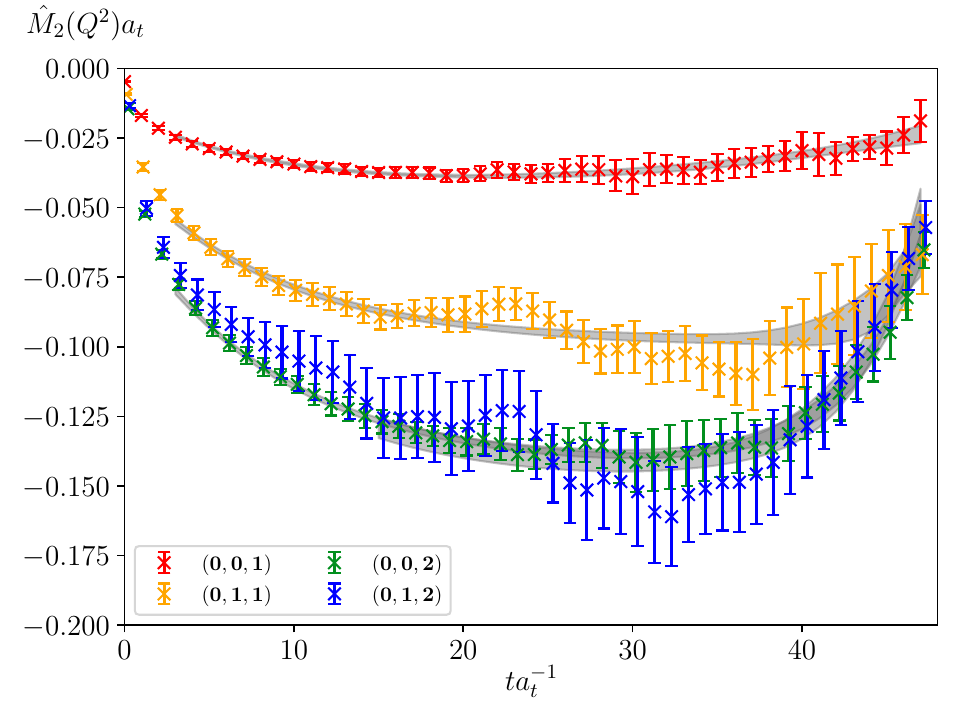}
  \includegraphics[width=5.5cm]{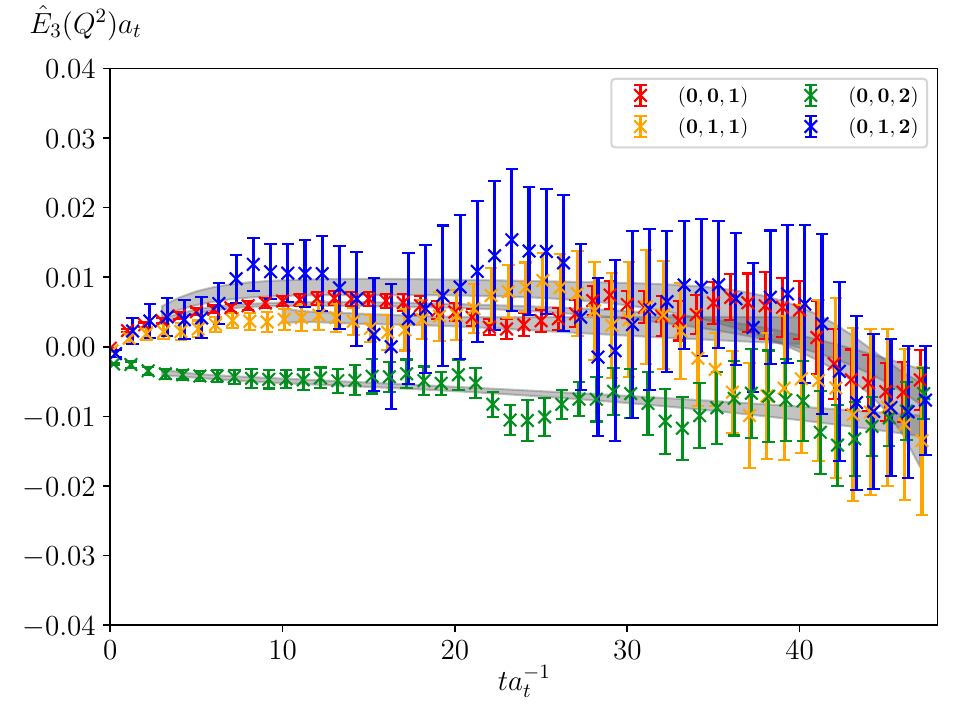}
\caption{Form Factors $\hat{E}_1(Q^2)$, $\hat{M}_2(Q^2)$ and $\hat{E}_3(Q^2)$ versus $ta_t^{-1}$ for the radiative transition $\chi_{c2}\rightarrow{J/\psi\gamma}$.
 The momentum of final particle is $\bf{p_f} = \frac{2\pi}{a_sL}(n_x\ n_y\ n_z)$, the legend denotes explicit value of $(n_x,n_y,n_z)$.
The points are lattice data and the shaded bands are the fit results using the function form in Eq.~(\ref{eq:tdependence2}).}
\label{fig:form factor21}
\end{figure*}

The decay width for $\chi_{c2}\rightarrow{J/\psi}\gamma$ involves only three on-shell form factors, namely, $E_1(0)$, $M_2(0)$ and $E_3(0)$, by the formula
\beq
\Gamma(\chi_{c2}&\to& \gamma J/\psi)=\frac{16}{45}\alpha\frac{|q|}{m_{\chi_{c2}}^2}\nonumber\\
&&\times \left(|\hat{E}_1(0)|^2+|\hat{M}_2(0)|^2+|\hat{E}_3(0)|^2\right),
\label{decay width}
\eeq
So we focus on the extraction of these three form factors at different $Q^2$ and then perform the interpolation (extrapolation) to get the on-shell values.
The procedure is very similar to that of $V(Q^2)$ for $J/\psi\to \gamma \eta_c$ except that $t_f$ is $48a_t$ in stead of $t_f=64a_t$ (We also calculate the three-point function with $t_f=64a_t$ and find the signals are very bad). The $t$-dependence of $\hat{E}_1(Q^2;t_f,t)$, $\hat{M}_2(Q^2;t_f,t)$ and $\hat{E}_3(Q^2;t_f,t)$ are shown in Fig.\ref{fig:form factor21}. By fitting these quantities using the function similar to Eq.~(\ref{eq:tdependence}), namely,
\beq\label{eq:tdependence2}
\hat{F}_k(Q^2,t)&=&\hat{F}_k(Q^2) \left(1+\delta_1^{(k)}(Q^2) e^{\Delta_1^{(k)}t}\right.\nonumber\\
 &&\left. +\delta_2^{(k)}(Q^2)e^{-\Delta_2^{(k)}(t_f-t)}\right),
\eeq
where $\hat{F}_k$ with $k=1,2,3$ refer to $\hat{E}_1$, $\hat{M}_2$ and $\hat{E}_3$, respectively. The form factors $\hat{F}_k(Q^2)$ at different $Q^2$ are listed in Table~\ref{tab:J2pp_formfactor}. For $\hat{E}_1(Q^2)$ and $\hat{M}_2(Q^2)$, the on-shell values $\hat{F}_k(Q^2=0)$ are interpolated through the function form
\beq
\hat{F}_k(Q^2)=\hat{F}_k(0)(1+\lambda_k Q^2)\exp(-\frac{Q^2}{16\beta_k^2}).
\label{eq:fitted formula for form factor}
\eeq

\begin{figure}[t]
  {\includegraphics[width=1.0\linewidth]{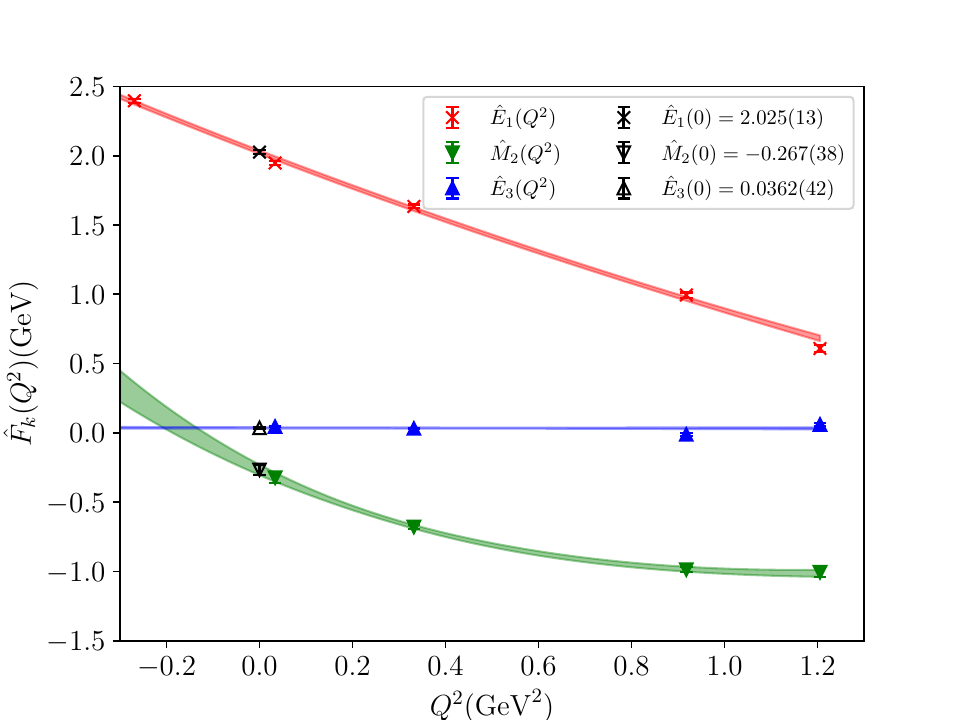}}
\caption{The $Q^2$ interpolation or extrapolation of $\hat{E}_1(Q^2)$, $\hat{M}_2(Q^2)$ and $\hat{E}_3(Q^2)$ for $\chi_{c2}\to\gamma J/\psi$. $\hat{E}_1(Q^2)$, $\hat{M}_2(Q^2)$ are fitted using Eq.~(\ref{eq:fitted formula for form factor}), while $\hat{E}_3(Q^2)$ is fitted using a linear equation in $Q^2$. The shaded bands illustrates the fit results and the black points are the values at $Q^2=0$.}
\label{fig:J2pp fit form factor1}
\end{figure}

Since the values of $\hat{E}_3(Q^2)$ are very small, we use a linear function $\hat{E}_3(Q^2)=\hat{E}_3(0)+a~Q^2$ to perform the extrapolation.
The fits for three form factors are illustrated in Fig.~\ref{fig:J2pp fit form factor1} by shaded bands. The extrapolated values of $\hat{F}_k(0)$ are also listed in Table~\ref{tab:J2pp_formfactor}.

Obviously, the electric dipole ($E_1$) contribution dominates the decay process $\chi_{c2}\to \gamma J/\psi$, and the hierarchy
$|E_1(0)|>|M_2(0)|\gg |E_3(0)|$ is described by the two ratios
\begin{eqnarray}
    a_2&=&\frac{M_2(0)}{\sqrt{E_1(0)^2+M_2(0)^2+E_3(0)^2}}=-0.130(18)\nonumber\\
    a_3&=&\frac{E_3(0)}{\sqrt{E_1(0)^2+M_2(0)^2+E_3(0)^2}}=0.0177(21),\nonumber\\
\end{eqnarray}
which are in agreement with the PDG values $a_2=-0.11(1)$ and $a_3=-0.003(10)$~\cite{ParticleDataGroup:2022pth}.

\begin{table}
 \centering \caption{The explicit value of the form factors $\hat{E_1}(Q^2)$, $\hat{M_2}(Q^2)$, and $\hat{E_3}(Q^2)$ for radiative transition $\chi_{c2}\rightarrow{J/\psi\gamma}$. The values are in physical units and are converted by $a_t^{-1}=6.894(31)~\text{GeV}$.}
 \label{tab:J2pp_formfactor}
 \begin{ruledtabular}
 \begin{tabular}{ccccc}
$\mathbf{n}$   &$Q^2$                &$\hat{E_1}(Q^2)$   &$\hat{M_2}(Q^2)$   &$\hat{E_3}(Q^2)$\\
              & ($\text{GeV}^2$)    &  ($\text{GeV})$   & ($\text{GeV})$    & ($\text{GeV})$\\
\hline
(0,1,2)       &1.21                 &0.609(28)         &-1.006(31)         &0.0608(92)     \\
(0,0,2)       &0.91                 &0.995(18)         &-0.986(20)         &-0.011(12)    \\
(0,1,1)       &0.33                 &1.634(14)         &-0.679(14)         &0.0321(51)    \\
(0,0,1)       &0.033                &1.948(15)         &-0.324(40)         &0.0459(51)    \\
(0,0,0)       &-0.27                &2.396(14)         &     -             & -             \\
\hline
-             & 0                   &2.025(13)          &-0.267(38)         &0.0362(42)
 \end{tabular}
 \end{ruledtabular}
 \end{table}

With the interpolated values of $\hat{F}_k(0)$ and the experimental value of the masses of the mesons involved, the partial decay width of the decay $\chi_{c2}$
is predicted to be
\beq
\Gamma(\chi_{c2}\to \gamma J/\psi)=368(5)~\text{keV},
\eeq
which can be compared with the PDG average of $374(10)~\text{keV}$~\cite{ParticleDataGroup:2022pth} as well as the previous lattice results of $361(9)~\text{keV}$~\cite{Yang:2012mya} and $380(30)~\text{keV}$~\cite{Dudek:2009kk}. This comparison calibrates the
uncontrolled systematic uncertainties of our calculation to some extent.

\begin{figure}[t]
  {\includegraphics[width=7cm]{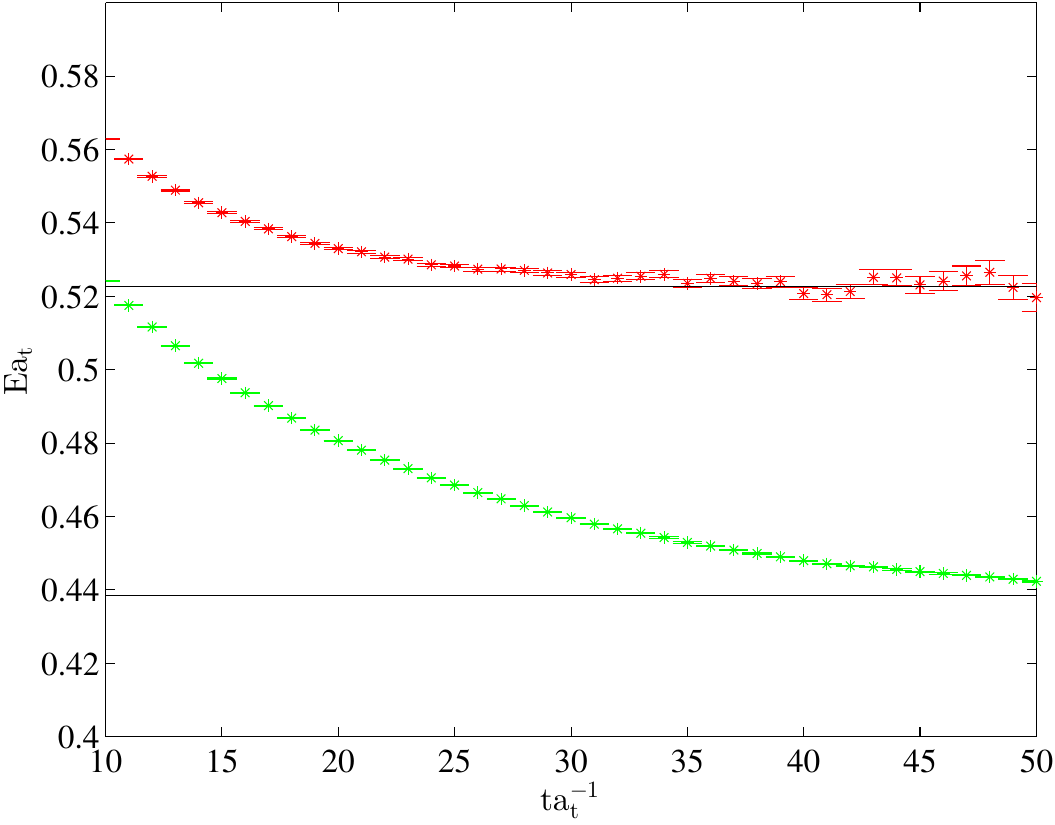}}
\caption{The effective energies of the correlation functions of $\mathcal{O}_{5i}^\mathbf{p}(t)$ (green) and $\mathcal{O}_{5i}^\text{op}(t,\mathbf{p})$ (red) for $\mathbf{p}$ mode $\mathbf{n}=(0,0,1)$. The black lines are shows the values of $E_{\eta_c}(\mathbf{p}a_t)$ and $E_{\chi_{c1}}(\mathbf{p})a_t$, respectively. Obviously, the optimized operator $\mathcal{O}_{5i}^\text{op}(t,\mathbf{p})$ couples to $\chi_{c1}$ when $t$ is large.}
\label{fig:varation_comparation}
\end{figure}

\subsection{$\psi_2\rightarrow\chi_{c1}\gamma$}
\label{resluts3}
The multipole decomposition of the transition matrix elements for $\psi_2\to \gamma \chi_{c1}$ ($2^{--}\to 1^{++}$) is exactly the same as that for $\chi_{c2}\to \gamma J/\psi$ ($2^{++}\to 1^{--}$) (see Eq.~(\ref{eq:2pp-1mm}))~\cite{Dudek:2009kk,Yang:2012mya}. The calculation of the related three-point function in Eq.~(\ref{eq:three-point}) is performed in the rest frame of the initial state $\psi_2$. The subtlety in this case is that, the generic quark bilinear operator $\mathcal{O}_{5i}\sim\bar{\psi}\gamma_5\gamma_i \psi$ for $\chi_{c1}$ couples both pseudoscalar mesons and axial vector mesons with the overlapping factors
\beq
\langle\Omega|\bar{\psi}\gamma_5\gamma_i\psi(0)|0^{-+}(\mathbf{p})\rangle&=&Z_P p_i\nonumber\\
\langle\Omega|\bar{\psi}\gamma_5\gamma_i\psi(0)|1^{++}(\mathbf{p},r)\rangle&=&Z_A \epsilon_i^{(r)}(\mathbf{p}),
\eeq
when the spatial momentum $\mathbf{p}$ is nonzero. Therefore, the contribution from pseudoscalar mesons (mainly $\eta_c$) should be eliminated when the three point function $G_{f\mu i}(t_f,t;\mathbf{p}_f, \mathbf{p}_i)$ in Eq.~(\ref{eq:three-point}) is computed. This is accomplished by choosing an optimized operator that couples predominantly to axial vector mesons. In doing so, for index $i$ of $\mathcal{O}_{5i}$ and each momentum $\mathbf{p}$ with $p_i\ne 0$, we adopt the momentum projected operator $\mathcal{O}_{5}^{\mathbf{p}}$ to calculate the correlation matrix
\beq
\mathbf{C}(t)=\left(\begin{array}{cc}
    \langle\Omega|\mathcal{O}_5^{\mathbf{p}}(t)\mathcal{O}_5^{\mathbf{p},\dagger}(0)|\Omega\rangle &
    \langle\Omega|\mathcal{O}_{5i}^{\mathbf{p}}(t)\mathcal{O}_5^{\mathbf{p},\dagger}(0)|\Omega\rangle\\
    \langle\Omega|\mathcal{O}_{5}^{\mathbf{p}}(t)\mathcal{O}_{5i}^{\mathbf{p},\dagger}(0)|\Omega\rangle &
    \langle\Omega|\mathcal{O}_{5i}^{\mathbf{p}}(t)\mathcal{O}_{5i}^{\mathbf{p},\dagger}(0)|\Omega\rangle
\end{array}\right).
\eeq
Since $\mathcal{O}_5^{\mathbf{p}}$ couples exclusively to pseudoscalar mesons ($\eta_c$ and its excited states), for properly chosen $t$ and $t_0$, by solving the generalized eigenvalue problem
$\mathbf{C}(t)\mathbf{v}=\lambda(t-t_0)\mathbf{C}(t_0)\mathbf{v}$ with $\mathbf{v}^T=(v_1,v_2)$ being an eigenvector, we can obtain the optimized operator that
couples to axial vector mesons as follows,
\beq
\mathcal{O}_{5i}^\text{op}(t,\mathbf{p})=v_1 \mathcal{O}_{5}^{\mathbf{p}}(t)+v_2 \mathcal{O}_{5i}^{\mathbf{p}}(t).
\eeq

The effectiveness of this prescription is illustrated by Figure~\ref{fig:varation_comparation}, where the effective energies are plotted for the correlation function of $\mathcal{O}_{5i}$ (in green) and that of the optimized operator $\mathcal{O}_{5i}^\text{op}$ (in red) for the momentum mode $\mathbf{n}=(0,0,1)$. It is seen that,
the effective energy of the former does not show a plateau but tends to the energy of $\eta_c$ when $t$ increases, while the effective energy of the latter reaches a plateau of a value consistent with the energy of $\chi_{c1}$ at this momentum. 
Therefore, for each momentum $\mathbf{p}$ mode of the final state $\chi_{c1}$, we use the optimized operator $\mathcal{O}_{5i}^\text{op}$ to calculated the three-point function $G_{f\mu i}$ in Eq.~(\ref{eq:three-point}). The related transition matrix elements are extracted similarly to the cases of $J/\psi\to\gamma\eta_c$ and $\chi_{c2}\to \gamma J/\psi$ through Eq.~(\ref{eq:ratio}).

\begin{figure*}[!htbp]
  {\includegraphics[width=5.5cm]{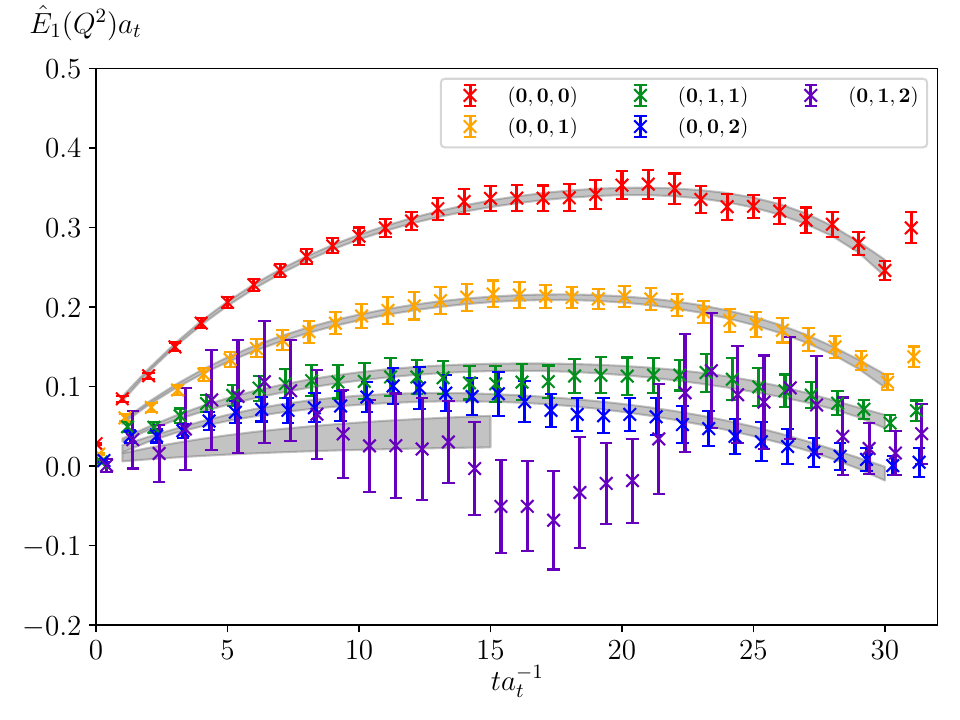}}
  {\includegraphics[width=5.5cm]{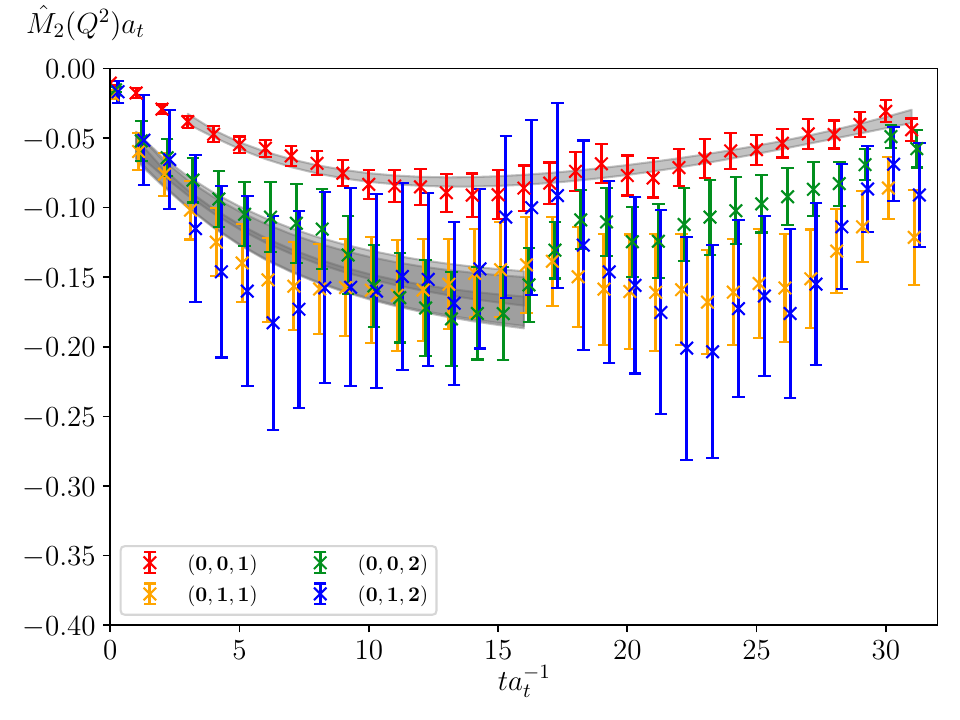}}
  {\includegraphics[width=5.5cm]{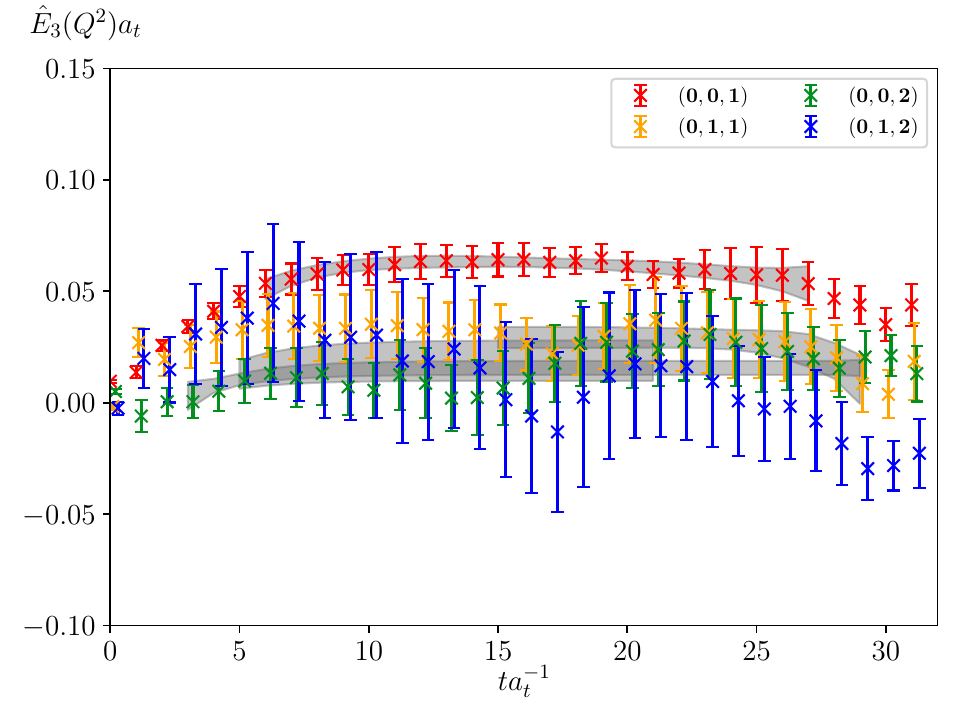}}
\caption{Form factors $\hat{E}_1(Q^2)$, $\hat{M}_2(Q^2)$ and $\hat{E}_3(Q^2)$ versus $ta_t^{-1}$ for the radiative transition $\psi_2\rightarrow\chi_{c1}\gamma$.  The momentum of final particle is $\bf{p_f} = \frac{2\pi}{a_sL}(n_x\ n_y\ n_z)$, the legend denotes explicit value of $(n_x,n_y,n_z)$. The points are lattice data and the shaded bands are the fit results using the function form in Eq.~(\ref{eq:tdependence2}).}
\label{fig: form factor31}
\end{figure*}

The electromagnetic multipole decomposition of the matrix element is exactly the same as that for $\chi_{c2}\to\gamma J/\psi$, and is expressed in terms of five form factors $E_1(Q^2)$,$M_2(Q^2)$, $E_3(Q^2)$, $C_1(Q^2)$ and $C_2(Q^2)$. Considering that the final state photon is transversely polarized, only the former three form factors contribute to the partial width of the process $\psi_2\to \gamma \chi_{c2}$, namely,
\beq
\Gamma(\psi_{2}&\to& \gamma \chi_{c1})=\alpha\frac{16}{45}\frac{|\mathbf{q}|}{m_{\psi_{2}}^2}\nonumber\\
&&\times \left(|\hat{E}_1(0)|^2+|\hat{M}_2(0)|^2+|\hat{E}_3(0)|^2\right).
\label{eq:width2--}
\eeq
For different values of $Q^2$, the three form factors $\hat{E}_1(Q^2)$, $\hat{M}_2(Q^2)$ and $\hat{E}_3(Q^2)$ are extracted similarly to the case of $\chi_{c2}\to \gamma J/\psi$, as shown in Fig.~\ref{fig: form factor31}, where the shaded bands illustrate the fit results using Eq.~(\ref{eq:tdependence}).

The final values of form factors $\hat{E}_1(Q^2)$, $\hat{M}_2(Q^2)$, $\hat{E}_3(Q^2)$ are listed in Table~\ref{tab:J2mm_formfactor} along with the extrapolated values at $Q^2=0$ using Eq.~(\ref{eq:fitted formula for form factor}). After putting the values of $F_k(Q^2=0)$ into Eq.~(\ref{eq:width2--}), the partial decay width of $\psi_2\to \gamma \chi_{c1}$ is predicted to be
\begin{equation}
    \Gamma(\psi_2\to \gamma \chi_{c1})=337(27)~\text{keV}.
\end{equation}
It is seen that, although the dominant electric dipole ($E_1$) contribution is similar to the case of $\chi_{c2}\to\gamma J/\psi$, the contributions from the magnetic quadrupole ($M_2$) and the electric octupole ($E_3$) are substantial. Accordingly, we give the predictions
\begin{eqnarray}
    a_2&=&\frac{M_2(0)}{\sqrt{E_1(0)^2+M_2(0)^2+E_3(0)^2}}=
    -0.485(37)\nonumber\\
    a_3&=&\frac{E_3(0)}{\sqrt{E_1(0)^2+M_2(0)^2+E_3(0)^2}}= 0.137(19).
\end{eqnarray}

\begin{figure}[t]
  {\includegraphics[width=8cm]{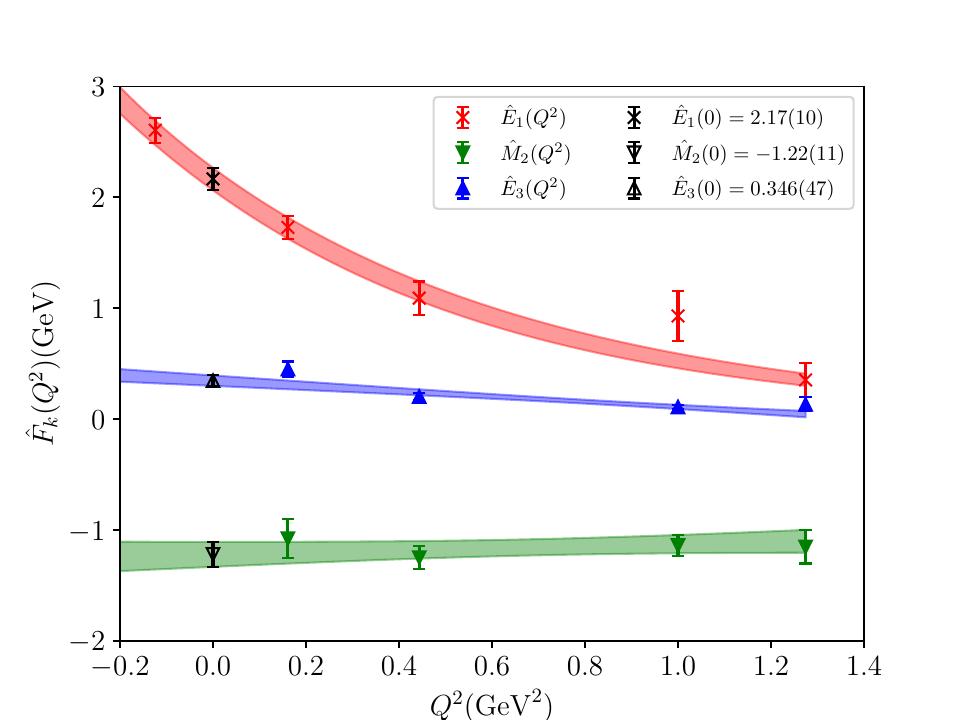}}
\caption{The $Q^2$ interpolation or extrapolation of $\hat{E}_1(Q^2)$, $\hat{M}_2(Q^2)$ and $\hat{E}_3(Q^2)$ for $\psi_2\to\gamma \chi_{c1}$. $\hat{E}_1(Q^2)$, $\hat{M}_2(Q^2)$ are fitted using Eq.~(\ref{eq:fitted formula for form factor}), while $\hat{E}_3(Q^2)$ is fitted using a linear equation in $Q^2$. The shaded bands illustrates the fit results and the black points are the values at $Q^2=0$.  }
\label{fig: J2mm fit form factor1}
\end{figure}

\begin{table}
 \centering \caption{The explicit value of form factor $\hat{E_1}(Q^2)$, $\hat{M_2}(Q^2)$, and $\hat{E_3}(Q^2)$ for
 radiative transition $\psi_2\rightarrow\chi_{c1}\gamma$.}
 \label{tab:J2mm_formfactor}
 \begin{ruledtabular}
 \begin{tabular}{ccccc}
$\mathbf{n}$   &$Q^2$         &$\hat{E_1}(Q^2)$   &$\hat{M_2}(Q^2)$   &$\hat{E_3}(Q^2)$\\
             & ($\text{GeV}^2$)    &  ($\text{GeV})$   & ($\text{GeV})$    & ($\text{GeV})$\\
\hline
(0,1,2)       &1.27                 &0.35(15)         &-1.15(15)          &0.133(62)\\
(0,0,2)       &1.00                 &0.92(22)         &-1.140(95)         &0.108(22)\\
(0,1,1)       &0.44                 &1.09(15)         &-1.24(10)          &0.203(33)\\
(0,0,1)       &0.16                 &1.72(10)         &-1.07(18)          &0.451(69)\\
(0,0,0)       &-0.12                &2.60(11)         &                   &\\
\hline
-             & 0                   &2.17(10)         &-1.22(11)          &0.346(47)\\
 \end{tabular}
  \end{ruledtabular}
 \end{table}

\begin{table}[t]
    \centering
    \caption{Comparison of the predictions of $\Gamma(\chi_{c2}\to\gamma J/\psi)$ and $\Gamma(\psi_2\to \gamma\chi_{c1})$ by different theoretical formalism. The abbreviations in the rightest column refer to the non-relativistic QCD approach (NRQCD), the non-relativistic quark models (QM), the relativistic quark models (RQM), the Bethe-Salpeter equation (BS), and the quenched lattice QCD (QLQCD) calculations, respectively. Various confining potentials are adopted in QM and RQM approaches, and the details can be found in the corresponding references. The experimental value and the predictions of this work are shown in bold numbers.
    \label{tab:pheno-prediction}}
    \begin{ruledtabular}
    \begin{tabular}{lll}
        $\Gamma(\chi_{c2}\to \gamma J/\psi)(\text{keV})$                & $\Gamma(\psi_2\to \gamma\chi_{c1})(\text{keV})$   & Formalism \\\hline
        $282$~\cite{Chao:1992hd}                                        & $250$~\cite{Qiao:1996ve}                          & NRQCD\\
        $401$~\cite{Chao:1992hd}                                        &  -                                                & NRQCD\\
        &&\\
        $315$                                                           & $260$                                             & QM ~\cite{Eichten:2002qv}\\
        $424$                                                           & $307$                                             & QM ~\cite{Barnes:2005pb}\\
        $473$                                                           & $342$                                             & QM ~\cite{Li:2009zu}\\
        $309$                                                           & $208$                                             & QM ~\cite{Li:2009zu}\\
        $327$                                                           & $281$                                             & QM ~\cite{Deng:2016stx}\\
        $338$                                                           & $291$                                             & QM ~\cite{Deng:2016stx}\\

        &&\\
        $313$                                                           & $268$                                             & RQM ~\cite{Barnes:2005pb}\\
        $448$                                                           & $297$                                             & RQM ~\cite{Ebert:2002pp}\\
        $309$                                                           & $215$                                             & RQM ~\cite{Ebert:2002pp}\\
        $292$                                                           & $215$                                             & RQM ~\cite{Ebert:2002pp}\\
        $327$                                                           & $215$                                             & RQM ~\cite{Ebert:2002pp}\\
        &&\\
          -                                                             & $265$                                             & BS~\cite{Li:2022qhg}\\
        &&\\
        $361(9)$                                                        &  -                                                & QLQCD~\cite{Yang:2012mya}\\
        $380(50)$                                                       &  -                                                & QLQCD~\cite{Dudek:2009kk}\\
        &&\\
        $\mathbf{368(5)}$                                                        & $\mathbf{337(27)}$                                         & This work\\
        &&\\
        $\mathbf{374(10)}$                                                       & -                                                 & PDG2022~\cite{ParticleDataGroup:2022pth}
    \end{tabular}
    \end{ruledtabular}
\end{table}

\subsection{Discussion}
As has been shown in the previous sections, the obtained form factors for transitions $J/\psi\to \gamma \eta_c$ and $\chi_{c2}\to \gamma J/\psi$ based on our lattice setup are consistent with previous lattice results. Especially, our prediction for the partial width and the hierarchy of $|E_1(0)|>|M_2(0)|\gg |E_3(0)|$ of the process $\chi_{c2}\to \gamma J/\psi$ are in quantitatively agreement with the experimental data. This comparison justifies the reliability of our predictions for the process $\psi_2\to \gamma \chi_{c1}$.

There have been quite a lot of phenomenological studies on radiative charmonium transitions using various theoretical frameworks, such as the non-relativistic QCD approach (NRQCD), the non-relativistic quark models (QM) with different confining potentials, the relativistic quark models and the Bethe-Salpeter wave function method, etc. Their predictions for the partial decay widths of $\chi_{c2}\to \gamma J/\psi$ and $\psi_2\to \gamma \chi_{c1}$ are collected in Table~\ref{tab:pheno-prediction} along with the precise references. Also shown are the previous lattice QCD predictions in the quenched approximation (QLQCD), the experimental values and the results in this work.

As far as the $\chi_{c2}\to \gamma J/\psi$ transition is concerned, the phenomenological predictions of the partial width range from 280 keV to 450 keV and are consistent with the experimental value $374(10)$ keV when considering the theoretical uncertainties owing to the model assumptions. The values by QLQCD are more converged and agree quantitatively with the PDG value. Our result $\Gamma(\chi_{c2}\to\gamma J/\psi)=368(5)~\text{keV}$ is the first prediction from the lattice QCD with light dynamical quarks, and is in excellent agreement with QLQCD results and the PDG value.

\begin{table}[t]
    \centering
    \caption{The branching-fraction ratios $\frac{\mathcal{B}(\psi_2(3823)\to X)}{\mathcal{B}(\psi_2(3823)\to\gamma \chi_{c1})}$ measured by BESIII~\cite{BESIII:2021qmo}
     with X referring to the decay channels $\gamma\chi_{c2}$, $\pi^+\pi^- J/\psi$, $\pi^0\pi^0 J/\psi$, $\eta J/\psi$, $\pi^0 J/\psi$ and $\gamma \chi_{c0}$.\label{tab:decay-channels}}
    \begin{ruledtabular}
    \begin{tabular}{cc}
       Channel ($X$)                & $\frac{\mathcal{B}(\psi_2(3823)\to X)}{\mathcal{B}(\psi_2(3823)\to\gamma \chi_{c1})}$\\
       \hline
       $\gamma\chi_{c2}$            & $0.28_{-0.11}^{+0.14}\pm 0.02$\\
       $\pi^+\pi^- J/\psi$          & $<0.06$\\
       $\pi^0\pi^0 J/\psi$          & $<0.11$\\
       $\eta J/\psi$                & $<0.14$\\
       $\pi^0 J/\psi$               & $<0.03$\\
       $\gamma \chi_{c0}$           & $<0.24$
    \end{tabular}
    \end{ruledtabular}
\end{table}
We also give the first lattice QCD prediction $\Gamma(\psi_2\to \gamma \chi_{c2})=337(27)~\text{keV}$, whose central value is slightly larger than
the phenomenological predictions (see Table~\ref{tab:pheno-prediction}), most of which are below 300 keV. On the other hand, BESIII measured the branching-fraction ratios $\mathcal{B}(\psi_2(3823)\to X)/\mathcal{B}(\psi_2(3823)\to\gamma \chi_{c1})$
with X referring to the decay channels $\gamma\chi_{c2}$, $\pi^+\pi^- J/\psi$, $\pi^0\pi^0 J/\psi$, $\eta J/\psi$, $\pi^0 J/\psi$ and $\gamma \chi_{c0}$~\cite{BESIII:2021qmo}, which are quoted in Table~\ref{tab:decay-channels}. These ratios are equivalently the ratios of the corresponding partial decay widths $\Gamma(\psi_2(3823)\to X)/\Gamma(\psi_2(3823)\to \gamma\chi_{c1})$. Based on these results, we can estimate the total width of $\psi_{2}$ as follows:
\begin{itemize}
    \item $\Gamma(\psi_2(3823)\to\gamma \chi_{c2})$: According to the branching-fraction ratio measured by BESIII, this partial width is estimated to be $94_{-39}^{+49}~\text{keV}$.
    \item $\Gamma(\psi_2(3823)\to \pi\pi J/\psi$): Although BESIII gives individual upper limits for the branching-fraction ratios $0.06$ and $0.11$ for $\pi^+\pi^- J/\psi$ and $\pi^0\pi^0 J/\psi$ decay channels, respectively, the isospin symmetry implies that $\Gamma(\pi^+\pi^- J/\psi)/\Gamma(\pi^0\pi^0 J/\psi)\approx 2$. Therefore, we assume $\Gamma(\psi_2(3823)\to \pi\pi J/\psi)/\Gamma(\psi_2(3823)\to \gamma \chi_{c1})<0.1$, which implies $\Gamma(\psi_2(3823)\to \pi\pi J/\psi)<34(3)~\text{keV}$. This is compatible with the QM prediction $\Gamma(\psi_2(3823)\to \pi\pi J/\psi)\approx 45~\text{keV}$~\cite{Eichten:2002qv}, but much smaller than the value of 160 keV predicted by Ref.~\cite{Wang:2015xsa}.
    \item $\Gamma(\psi_2(3823)\to \eta J/\psi)$: The flavor SU(3) symmetry requires the $\eta$ in the final state is produced through gluons coupling to its flavor singlet component. The small $\eta-\eta'$ mixing angle $\theta$ (the partial width is proportional to $\sin^2 \theta$) and the centrifugal barrier ($\eta$ and $J/\psi$ are in $P$-wave) suppresses the decay rate of this process, but the QCD $U_A(1)$ anomaly enhances the coupling of gluons to $\eta$ and may counteract the suppression. Referring to the branching fraction ratio $\frac{\mathcal{B}(\psi(3770)\to \pi\pi J/\psi)}{\mathcal{B}(\psi(3770)\to \eta J/\psi)}\approx 3$~\cite{ParticleDataGroup:2022pth}, $\Gamma(\psi_2(3823)\to \eta J/\psi)< 20~\text{keV}$ might be a reasonable estimate even though BESIII gives a higher upper limit.
    \item $\Gamma(\psi_2(3823)\to\gamma \chi_{c0}, \gamma \eta_c)$: These two partial widths are predicted to be $\sim 1$ keV by a phenomenological study through the Bethe-Salpeter equation approach~\cite{Li:2022qhg}.
    \item $\Gamma(\psi_2(3823)\to \pi^0 J/\psi)$: The partial width of this isospin breaking decay channel can be neglected.
    \item $\Gamma(\psi_2(3823)\to \text{light hadrons})$: The total decay widths of $\psi_2(3823)\to \text{light hadrons}$ can be approximated by $\Gamma(\psi_2\to ggg)\sim 36~\text{keV}$~\cite{Eichten:2002qv}.
\end{itemize}
Summing over all the contributions mentioned above, we can give a raw estimate of the total width of $\psi_2(3823)$
\begin{equation}
    \Gamma(\psi_2(3823))\approx 520\pm 100 ~\text{keV},
\end{equation}
where the uncertainty mainly comes from the partial widths of $\psi_2(3823)\to \gamma \chi_{c1}, \chi_{c2}$, and can be reduced by a refined lattice
QCD calculation of $\Gamma(\psi_2(3823)\to \gamma \chi_{c1})$ and a direct lattice calculation of $\Gamma(\psi_2(3823)\to \gamma \chi_{c2})$ in the future.

\section{Summary}\label{Discussion}
We perform an exploratory $N_f=2$ lattice QCD study on the radiative transition $\psi_2(3823)\to \gamma \chi_{c1}$ in the framework of the distillation method. On a single gauge ensemble with a pion mass $m_\pi \sim 350~\text{MeV}$, the electromagnetic multipole form factors are extracted for the processes $J/\psi\to\gamma \eta_c$, $\chi_{c2}\to\gamma J/\psi$ and $\psi_2\to\gamma\chi_{c1}$. The obtained $\hat{V}(0)=2.083(11)$ for $J/\psi\to\gamma \eta_c$ is consistent with previous lattice results, but the result $\Gamma(J/\psi\to\gamma\eta_c)=2.77(3)~\text{keV}$ is still larger than the PDG average. For $\chi_{c2}\to\gamma J/\psi$, we extract the on-shell form factors $E_1(0)$, $M_2(0)$ and $E_3(0)$, whose hierarchy
$|E_1(0)|>|M_2(0)|\gg |E_3(0)|$ is in quantitative agreement with the experimental results. We predict $\Gamma(\chi_{c2}\to\gamma J/\psi)=368(5)~\text{keV}$, which is in excellent agreement with the PDG value $374(10)~\text{keV}$ and previous QLQCD results. This is the first result from lattice QCD with dynamical light quarks. No quenched effects are observed here.

We present the first lattice QCD prediction of the partial decay width $\Gamma(\psi_2(3823)\to \gamma \chi_{c1})=337(27)~\text{keV}$, whose central value is higher than most of the phenomenological results. According to the BESIII measurement of branching fractions of $\psi_2(3823)$ decay channels and some phenomenological results, we estimate the total width $\Gamma(\psi_2(3823))=520(100)~\text{keV}$. A direct lattice QCD calculation of the partial widths of $\psi_2\to \gamma \chi_{c2}, \gamma \chi_{c0}, \gamma \eta_c$ will reduce the uncertainty of the total width. This can be fulfilled in the future.

\section{Acknowledgments}

This work is supported in part by the
National Natural Science Foundation of China (NNSFC) under Grants No.~12075176, No.11935017, No. 12293060, No.12293065 and No.12070131001 (CRC 110 by DFG and NNSFC), the Innovation Capability Support Program of Shaanxi (Program No.~2022KJXX-42), and 2022 Shaanxi University Youth Innovation Team Project (K20220186). CY also acknowledges the support by the National Key Research and Development Program of China (No. 2020YFA0406400) and the Strategic Priority Research Program of Chinese Academy of Sciences (No. XDB34030302).
The Chroma software system~\cite{Edwards:2004sx} and QUDA library~\cite{Clark:2009wm,Babich:2011np} are acknowledged. The simulations were performed on the HPC clusters at Institute of High Energy Physics (Beijing) and China Spallation Neutron Source (Dongguan).

 \bibliographystyle{apsrev4-1}
\input Radiative_transitions.bbl

\end{document}